\documentclass[aps,preprint,nofootinbib]{revtex4}
\usepackage{graphicx}

\begin{document}
\def\dirac#1{#1\llap{/}}
\def\pv#1{\vec{#1}_\perp}

\newcommand{\slashl}[1]{\not{\!\!#1}}
\newcommand{\slashs}[1]{\not{\!#1}}

\title{B-Meson Wavefunction with Contributions from 3-particle Fock States}
\author{Tao Huang$^{1,2}$\footnote{email:
huangtao@mail.ihep.ac.cn}, Cong-Feng Qiao$^{1,3}$\footnote{email:
qiaocf@gucas.ac.cn}, Xing-Gang Wu$^{4}$\footnote{email:
wuxg@itp.ac.cn}}
\address{$^1$CCAST(World
Laboratory), P.O.Box 8730, Beijing 100080, P.R.China,\\
$^2$Institute of High Energy Physics, Chinese Academy of Sciences,
P.O.Box 918(4), Beijing 100049, P.R. China,\\
$^3$Department of Physics, Graduate School of the Chinese Academy
of Sciences, Beijing 100049, P.R. China\\
$^4$Institute of Theoretical Physics, Chinese Academy of Sciences,
P.O.Box 2735, Beijing 100080, P.R. China.}

\begin{abstract}
The B-meson light-cone wavefunctions, $\Psi_{\pm}(\omega,z^2)$, are
investigated up to the next-to-leading order in Fock state expansion
in the heavy quark limit. In order to know the transverse momentum
dependence of the B-meson wavefunctions with 3-particle Fock states'
contributions, we make use of the relations between 2- and 3-
particle wavefunctions derived from the QCD equations of motion and
the heavy quark symmetry, especially two constraints derived from
the gauge field equation of motion are employed. Our results show
that the use of gluon equation of motion can give a constraint on
the transverse momentum dependence $\chi(\omega,\mathbf{k}_\perp)$
of the B-meson wavefunctions, whose distribution tends to be a
hyperbola-like curve under the condition $0<c_1<1$, which is quite
different from the WW type wavefunctions, whose transverse momentum
dependence $\chi^{WW}(\omega,\mathbf{k}_\perp)$ is merely a delta
function. Based on the derived results, we propose a simple model
for the B-meson wavefunctions with 3-particle Fock
states' contributions.\\

\noindent {\bf PACS numbers:} 12.38.Aw, 12.39.Hg, 14.40.Nd

\noindent {\bf Keywords:} B-meson wavefunction, 3-particle Fock
state, heavy quark symmetry
\end{abstract}
\maketitle

\section{Introduction}

The non-perturbative light-cone (LC) wavefunction (WF)/distribtuion
amplitude (DA) of the B meson plays an important role for making
reliable predictions for exclusive B meson decays. The B-meson DA
has been investigated in various approaches
\cite{bdistribution1,bdistribution2,braun,lange,qiao0,beneke,
descotes,geyer,alex}. Recently, Ref.\cite{libwave} claims that it is
the B-meson WF rather than the B-meson DA that is more relevant to
the B decays, and in the framework of the $k_T$-factorization
theorem \cite{kt}, they proved that the B-meson WF is renormalizable
after taking into account the renormalization-group (RG) evolution
effects, meanwhile, the undesirable feature \cite{braun} of the
B-meson DA under evolution can be removed. Since the B-meson WF
still poses a major source of uncertainty in the study of the B
decays, hence, theoretically, it is an important issue to study on
it.

Ref. \cite{qiao}, as well \cite{bwave}, presents an analytic
solution for the B-meson WFs $\Psi_{\pm}(\omega,z^2)$, which
satisfies the constraints from the QCD equations of motion and the
heavy-quark symmetry \cite{heavyquark}. It is found that in the
Wandzura-Wilczek(WW) approximation \cite{ww}, which corresponds to
the valence quark distribution, the B-meson WFs can be determined
uniquely in terms of the ``effective mass'', the $\bar{\Lambda}$,
defined in the Heavy Quark Effective Theory (HQET) \cite{hqet}.
Ref.\cite{huangwu} shows that when taking $\bar{\Lambda}\in
(0.5GeV,0.6GeV)$, one can give a reasonable perturbative QCD result
for $B\to\pi$ transition form factor that is consistent with what
was obtained in the LC sum rule calculation \cite{sumrule} and the
lattice QCD simulation \cite{lattice}.

It should be noted that in the WW approximation, the obtained
analytic results for the B-meson WFs $\Psi_{\pm}(\omega,z^2)$ are
unique, and the only missing part for practical numerical use is the
RG evolution effect. However, there is very limited knowledge on the
higher Fock states' contributions. In Ref.\cite{qiao0}, the B-meson
distributions $\phi_{\pm}(\omega)$ with 3-particle Fock states are
given and a rough estimation presented there shows that the
3-particle Fock states' contributions might considerably broaden the
transverse momentum distribution that is derived from the WW
approximation. Recently, based on the QCD sum rule analysis and
taking only the two 3-particle distributions $\Psi_A(\rho,\xi)$ and
$\Psi_V(\rho,\xi)$ into consideration, Ref.\cite{alex} connects the
asymptotic behavior of their difference to the well-known
quark-antiquark-gluon DA $\varphi_{3\pi}(\alpha_i)$
($\alpha_i(i=1,2,3)$ are the fractions of the pion momentum carried
by the corresponding partons and satisfy
$\alpha_1+\alpha_2+\alpha_3=1$), i.e. in the small $\rho$ and $\xi$
region, $\Psi_V(\rho,\xi)-\Psi_A(\rho,\xi)\sim
\varphi_{3\pi}(\alpha_1,1-\alpha_1-
\alpha_3,\alpha_3)|_{\alpha_1=\rho/m_B,\alpha_3=\xi/m_B}\sim\rho\xi^2$.

In this paper, we are going to investigate the B-meson WFs
$\Psi_{\pm}(\omega,z^2)$ with the contributions from 3-particle Fock
states. From the heavy-quark symmetry and the equations of motion
for the light degrees of freedom, we get several constraints on the
behaviors of WFs. By adopting some assumptions, we will solve the
B-meson WFs approximately, especially, by taking the aforementioned
asymptotic behavior of the difference between $\Psi_A(\rho,\xi)$ and
$\Psi_V(\rho,\xi)$ into account, we try to derive an explicit form
for the B-meson distributions $\phi_{\pm}(\omega)$ that include the
3-particle Fock states' contributions.

The paper is organized as follows. In Sec.II, several differential
equations for the B-meson WFs are obtained by using the heavy quark
symmetry and the equations of motion for the light degrees of
freedom. In Sec.III, approximate solutions for the B-meson WFs
including the contributions from the 3-particle Fock states are
investigated under three assumptions. A new model for the B-meson
WFs and some discussions over its phenomenological implication are
presented in Sec.IV. The last section is reserved for a summary.

\section{Differential equations for the B-meson wavefunctions}

In HQET, the B-meson WFs $\tilde{\Psi}_{\pm}(t,z^2)$ can be
defined in terms of the vacuum-to-meson matrix element of the
nonlocal operators \cite{grozin}:
\begin{equation}\label{hqeteq}
\langle 0 | \bar{q}(z) \Gamma h_{v}(0) |\bar{B}(p) \rangle = -
\frac{i f_{B} M}{2} {\rm Tr} \Bigg[ \gamma_{5}\Gamma \frac{1 +
\slash\!\!\! v}{2} \!\!\!\times \Bigg\{ \tilde{\Psi}_{+}(t,z^2)-
\slash\!\!\! z \frac{\tilde{\Psi}_{+}(t,z^2)
 -\tilde{\Psi}_{-}(t,z^2)}{2t}\Bigg\} \Bigg],
\end{equation}
where $z^{\mu}=(0, z^{-}, \mathbf{z}_\perp)$, $z^{2}= -
\mathbf{z}_{\perp}^{2}$, $v^{2} = 1$, $t=v\cdot z$, and $p^{\mu} =
Mv^{\mu}$ is the 4-momentum of the B meson with mass $M$, $h_{v}(x)$
denotes the effective $b$-quark field and $\Gamma$ is a generic
Dirac matrix. The path-ordered gauge factors are implied between the
constituent fields. Note that in the above definition, the
separation between quark and antiquark is not restricted on the LC
($z^{2}=0$). For a fast moving meson, $t\to\infty$,
Eq.(\ref{hqeteq}) shows that $\tilde{\Psi}_{+}(t,z^2)$ is the
leading-twist WF and $\tilde{\Psi}_{-}(t,z^2)$ is the subleading
one. To know more about the twist structures for the B meson,
readers are recommended to refer to Ref.\cite{geyer} for details,
where the relation between the geometric twist and the dynamic twist
was discussed.

In the heavy quark limit, the general Lorentz decomposition of the
3-particle matrix elements can attribute to four independent
3-particle WFs similar to the 3-particle LC DAs \cite{qiao} , i.e.
$\tilde{\Psi}_V(t,u,z^2)$, $\tilde{\Psi}_A(t,u,z^2)$,
$\tilde{X}_A(t,u,z^2)$ and $\tilde{Y}_A(t,u,z^2)$:
\begin{eqnarray}
\lefteqn{\langle 0 | \bar{q} (z) \, g G_{\mu\nu} (uz)\, z^{\nu}
\, \Gamma \, h_{v} (0) | \bar{B}(p) \rangle}\nonumber \\
&=& \frac{1}{2}\, f_B M \, {\rm Tr}\, \left[ \, \gamma_5\, \Gamma
\, \frac{1 + \slashs{v}}{2}\, \biggl\{ ( v_{\mu}\slashs{z}
 - t \, \gamma_{\mu} )\  \left( \tilde{\Psi}_A (t,u,z^2)
- \tilde{\Psi}_V (t,u,z^2) \right) \right. \label{3elements}\\
& & - i \, \sigma_{\mu\nu} z^{\nu}\, \tilde{\Psi}_V (t,u,z^2) -
\left. \left(z_{\mu}-\frac{z^2 v_{\mu}}{t}\right) \tilde{X}_A
(t,u,z^2)\, + \left(\frac{z_{\mu}}{t}
\slashs{z}-\frac{z^2\gamma_{\mu}}{t}\right) \,\tilde{Y}_A
\,(t,u,z^2) \biggr\} \, \right]\ . \nonumber
\end{eqnarray}
Here, the $z^2$-dependent terms are kept explicitly in order to
get the transverse momentum dependence of the B-meson WFs.

By using the equation of motion for the light quark,
$\bar{q}\stackrel{\leftarrow}{\slashl{D}}=0$, and HQET equation of
motion for the heavy quark, $v \cdot D h_{v} = 0$, one can obtain
four independent differential equations, which correlate
$\tilde{\Psi}_{+}(t,z^2)$ and $\tilde{\Psi}_{-}(t,z^2)$ to the
3-particle WFs:
\begin{eqnarray}
\frac{\partial \tilde\Psi_{-}(t,z^2)}{\partial t}&-&
\frac{\tilde\Psi_+(t,z^2)-\tilde\Psi_-(t,z^2)}{t}-\frac{z^2}{t}
\frac{\partial}{\partial z^2}[\tilde\Psi_+(t,z^2)
-\tilde\Psi_-(t,z^2)]\nonumber\\
&=& \int_0^1 udu\; \left[2t[\tilde{\Psi}_A(t,u,z^2) - \tilde{\Psi}_V
(t,u,z^2)]+ 3\frac{z^2}{t}
\tilde{Y}_A(t,u,z^2)-\frac{z^2}{t}\tilde{X}_A(t,u,z^2)\right] ,
\label{original1} \\
\frac{\partial\tilde\Psi_+(t,z^2)}{\partial
t}&-&\frac{\partial\tilde\Psi_-(t,z^2)}{\partial
t}-\frac{\tilde\Psi_+(t,z^2)-\tilde\Psi_-(t,z^2)}{t}
+4t\frac{\partial\tilde\Psi_{+}(\omega,z^2)}{\partial
z^2}\nonumber\\
&=& \int_0^1 udu\; 2t[ \tilde{\Psi}_A (t,u,z^2)+ 2\, \tilde{\Psi}_V
(t,u,z^2) + \tilde{X}_A (t,u,z^2)] ,
\label{original2}\\
\frac{\partial\tilde{\Psi}_{+}(t,z^2)}{\partial t}&-&
\frac{1}{2t}\left[\tilde{\Psi}_{+}(t,z^2) - \tilde{\Psi}_{-}
(t,z^2)\right] + i \bar{\Lambda}\tilde{\Psi}_{+}(t,z^2)
+ 2t \frac{\partial \tilde{\Psi}_{+}(t,z^2)}{\partial z^{2}} \nonumber \\
&& \!\!\!\!\!\!\!\!\!\!\!\!\!\!\!\!\!\!=\int_0^1 du\,(u-1)\,\left[t[
\tilde{\Psi}_A (t,u,z^2) + \tilde{X}_A
(t,u,z^2)]-\frac{z^2}{t}\tilde{X}_A(t,u,z^2) +\frac{z^2}{t}
\tilde{Y}_A(t,u,z^2)\right] \label{original3}
\end{eqnarray}
and
\begin{eqnarray}
\frac{\tilde{\Psi}_{+}(t,z^2)}{\partial t}
&-&\frac{\tilde{\Psi}_{-}(t,z^2)}{\partial t} +\left(i \bar{\Lambda}
-\frac{1}{t}\right) \left[\tilde{\Psi}_{+}(t,z^2)-
\tilde{\Psi}_{-}(t,z^2)\right] + 2t \left[ \frac{\partial
\tilde{\Psi}_{+}(t,z^2)}{\partial z^{2}}-
\frac{\partial \tilde{\Psi}_{-}(t)}{\partial z^{2}}\right] \nonumber \\
&=&  2t \,\int_0^1 du\,(u-1)\, [ \tilde{\Psi}_A (t,u,z^2) +
\tilde{Y}_A (t,u,z^2) ] \ , \label{original4}
\end{eqnarray}
where $\bar{\Lambda} = M - m_{b} = \frac{iv\cdot \partial \langle 0|
\bar{q} \Gamma h_{v} |\bar{B}(p) \rangle} {\langle 0| \bar{q} \Gamma
h_{v} |\bar{B}(p) \rangle} $ is the usual ``effective mass'' of the
B meson in the HQET. By taking the LC limit, one may easily find
that our present results agree with those in Ref.\cite{qiao}. Making
the Fourier transformation, by virtue of the formulae given in the
appendix, we obtain,
\begin{eqnarray}\label{eq:1}
\omega \frac{\partial \Psi_{-}(\omega,z^2)}{\partial \omega} + z^2
\left[\frac{\partial \Psi_{+}(\omega,z^2)}{\partial z^2}
-\frac{\partial \Psi_{-}(\omega,z^2)}{\partial
z^2}\right]+\Psi_{+}(\omega,z^2) &=& I(\omega,z^2),\\
\label{eq:2} \left[\omega\frac{\partial}{\partial\omega}+2\right]
(\Psi_{+}(\omega,z^2)-\Psi_{-}(\omega,z^2))+
4\frac{\partial^2}{\partial\omega^2}
\frac{\partial\Psi_{+}(\omega,z^2)}{\partial z^2} &=&
J(\omega,z^2)\;,\\
\label{eq:3}
\left[(\omega-\bar\Lambda)\frac{\partial}{\partial\omega}+
\frac{3}{2}\right]\Psi_{+}(\omega,z^2)-\frac{1}{2}\Psi_{-}(\omega,z^2)+
2\frac{\partial^2}{\partial\omega^2}
\frac{\partial\Psi_{+}(\omega,z^2)}{\partial z^2} &=&
M(\omega,z^2)+N(\omega,z^2)
\end{eqnarray}
and
\begin{eqnarray}\label{eq:4}
\bar\Lambda[\Psi_{+}(\omega,z^2)-\Psi_{-}(\omega,z^2)]
+2\frac{\partial^{2}}{\partial \omega\partial z^2}[
\Psi_{+}(\omega,z^2)+\Psi_{-}(\omega,z^2)]&=& L(\omega,z^2)\;,
\end{eqnarray}
where the 3-particle source terms are
\begin{eqnarray}
I(\omega,z^2) &=& 2\frac{d}{d\omega}\int_{0}^{\omega}d\rho
\int_{\omega - \rho}^{\infty}\frac{d\xi}{\xi}
\frac{\partial}{\partial \xi}\left[ \Psi_{A}(\rho, \xi,z^2)-
\Psi_{V}(\rho, \xi,z^2)\right]\nonumber\\
&& -z^2\int_0^\omega d\rho\int_{\omega-\rho}^{\infty}
\frac{d\xi}{\xi}\left(\frac{\omega-\rho}{\xi}\right)
[X_A(\rho,\xi,z^2)-3Y_A(\rho,\xi,z^2)]\ ,\label{funI}\\
J(\omega,z^2) &=& 2\frac{d}{d\omega}\int_{0}^{\omega}d\rho
\int_{\omega - \rho}^{\infty}\frac{d\xi}{\xi}
\frac{\partial}{\partial \xi}\left[ \Psi_{A}(\rho, \xi,z^2)
+2\Psi_{V}(\rho, \xi,z^2)+X_A(\rho,\xi,z^2)\right]\ , \label{funK}\\
L(\omega,z^2)&=& -2\int_{0}^{\omega}d\rho \int_{\omega -
\rho}^{\infty}\frac{d\xi}{\xi}\frac{\partial}{\partial\xi}
[Y_A(\rho,\xi,z^2)-2\Psi_V(\rho,\xi,z^2)-X_{A}(\rho,
\xi,z^2)]\nonumber\\
&& +2 \frac{d}{d\omega}\int_{0}^{\omega}d\rho \int_{\omega
-\rho}^{\infty} \frac{d\xi}{\xi}[\Psi_A(\rho,
\xi,z^2)+Y_A(\rho,\xi,z^2)]\ ,\label{funL}\\
M(\omega,z^2)&=&\frac{d}{d\omega}\int_{0}^{\omega}d\rho
\int_{\omega - \rho}^{\infty}\frac{d\xi}{\xi}
\frac{\partial}{\partial \xi}\left[ \Psi_{A}(\rho, \xi,z^2)+
X_{A}(\rho, \xi,z^2)\right]\nonumber\\
&& -\frac{d^2}{d\omega^2}\int_{0}^{\omega}d\rho \int_{\omega
-\rho}^{\infty} \frac{d\xi}{\xi}[\Psi_A(\rho,
\xi,z^2)+X_A(\rho,\xi,z^2)] \label{funM}
\end{eqnarray}
and
\begin{eqnarray}
N(\omega,z^2)&=& z^2\int_0^\omega d\rho\int_{\omega-\rho}^{\infty}
\frac{d\xi}{\xi}\left(\frac{\omega-\rho}{\xi}\right)
[Y_A(\rho,\xi,z^2)-X_A(\rho,\xi,z^2)]\nonumber\\
&& -z^2\int_0^\omega d\rho\int_{\omega-\rho}^{\infty}
\frac{d\xi}{\xi}[Y_A(\rho,\xi,z^2)-X_A(\rho,\xi,z^2)]\ .
\label{funN}
\end{eqnarray}

Secondly, by using Eq.(\ref{3elements}) and the gluon equation of
motion, $D^{\mu} G^a_{\mu\nu}(z)=0$, whose source term that induces
even higher Fock state' contribution is neglected in this work, one
can obtain two more independent relations among the 3-particle WFs:
\begin{eqnarray}\label{fr1}
&&
\left[3\tilde{\Psi}_A(t,u,z^2)+4\tilde{Y}_A(t,u,z^2)-\tilde{X}_A(t,u,z^2)\right]
+t\frac{\partial}{\partial t} \left[\tilde{\Psi}_A(t,u,z^2)
+2\tilde{Y}_A(t,u,z^2)-\right.\nonumber\\
&& \left.\tilde{X}_A(t,u,z^2)\right]
+z^2\left[\frac{1}{t^2}\left[\tilde{Y}_A(t,u,z^2)
-\tilde{X}_A(t,u,z^2)\right]-\frac{1}{t}\frac{\partial}{\partial
t} \left[\tilde{Y}_A(t,u,z^2) -\tilde{X}_A(t,u,z^2)\right]
\right.\nonumber\\
&&+2\frac{\partial}{\partial z^2}[\tilde{\Psi}_A(t,u,z^2)\left.+
\tilde{Y}_A(t,u,z^2)]\right]=0
\end{eqnarray}
and
\begin{eqnarray}\label{fr2}
&&\frac{2}{t}[\tilde{X}_A(t,u,z^2)
-\tilde{Y}_A(t,u,z^2)]+2\frac{z^2}{t}\frac{\partial}{\partial
z^2}[\tilde{X}_A(t,u,z^2)-\tilde{Y}_A(t,u,z^2)]\nonumber\\
&&-2t\frac{\partial}{\partial z^2}\left[\tilde{\Psi}_A(t,u,z^2)+
\tilde{X}_A(t,u,z^2)\right]-\frac{\partial}{\partial t}
\left[\tilde{\Psi}_A(t,u,z^2)+\tilde{Y}_A(t,u,z^2)\right]=0\ .
\end{eqnarray}
Taking the LC limit in Eq.(\ref{fr1}), we get
\begin{equation}
\left[3\tilde{\Psi}_A(t,u)+4\tilde{Y}_A(t,u)-\tilde{X}_A(t,u)\right]
+t\frac{\partial}{\partial t} \left[\tilde{\Psi}_A(t,u)
+2\tilde{Y}_A(t,u)-\tilde{X}_A(t,u)\right]=0\ ,
\end{equation}
with $F(t, u)=F(t,u,z^2)|_{z^2\to 0}$ ($F=\{ \tilde\Psi_{V},
\tilde\Psi_{A}, \tilde X_{A}\}$). By doing the Fourier
transformation and exploiting the boundary conditions,
$F(\rho,\xi)|_{\rho\to\infty,\xi\to\infty}\to 0$ (F=$\Psi_A$,
$X_A$, $Y_A$), we obtain a relation among the double moments of
the 3-particle DAs,
\begin{equation}\label{momentrelation2}
3\left[\Psi_A(\rho,\xi)\right]_{1}^{1}+
4\left[Y_{A}(\rho,\xi)\right]_{1}^{1}=\left[X_{A}(\rho,\xi)
\right]_{1}^{1},
\end{equation}
where $\left[F\right]^{i}_{j}$ are double moments of the 3-particle
distributions,
\begin{displaymath}
\left[F\right]^{i}_{j} =\int_{0}^{\infty}d\rho \int_{0}^{\infty}d\xi
\ \rho^{i-j} \xi^{j-1} F(\rho, \xi)\ ,\qquad \left(F=\{ \Psi_{V},
\Psi_{A}, X_{A}\}\right).
\end{displaymath}

\section{Approximate solution for the B-meson WFs
with 3-particle Fock states}

In the following, we shall give an approximate solution for the
B-meson WFs with 3-particle Fock states' contributions by solving
the differential equations as shown in Sec.II. Before doing this,
as a basis and to be self-consistent, we first recollect the
results in the WW approximation (i.e.
$I(\omega)=J(\omega)=L(\omega)= M(\omega)=N(\omega)=0$), then,
derive several constraints for the B-meson DAs
$\phi_{\pm}(\omega)$, where the B-meson DAs can be obtained by
taking the LC limit of the B-meson WFs, i.e.
$\phi_\pm(\omega)\equiv \lim_{z^2\to 0}\Psi_{\pm}(\omega,z^2)$.

\subsection{B-meson WFs $\Psi^{WW}_{\pm}(\omega,z^2)$ in the WW approximation}
\label{suba}

When ignoring the 3-particle Fock states' contributions, i.e.
setting $I(\omega)=J(\omega)=L(\omega)=M(\omega)=N(\omega)=0$, one
can readily obtain the B-meson WW-type WFs
$\Psi^{WW}_{\pm}(\omega,z^2)$. In this case, the two DAs
$\phi^{WW}_{\pm}(\omega)$ take the form\cite{bwave,qiao0}:
\begin{equation}\label{phiww}
\phi^{WW}_{+}(\omega)=\frac{\omega}{2\bar\Lambda^2}\theta(\omega)
\theta(2\bar\Lambda-\omega)
 ; \;\;\;
\phi^{WW}_{-}(\omega)=\frac{(2\bar\Lambda-\omega)}{2\bar\Lambda^2}
\theta(\omega) \theta(2\bar\Lambda-\omega) ,
\end{equation}
and the transverse part, $\chi^{WW}(\omega,z^2)$, is a zero-{\it
th} normal Bessel function. $\theta(x)$ is the usual unit step
function, which equals to $0$ for $x<0$ and $1$ for $x\geq 0$.
Taking the Fourier transformation,
$\tilde\Psi_{\pm}(\omega,\mathbf{k}_\perp)=\int
d^2\mathbf{z}_{\perp} \exp(-i \mathbf{k}_\perp\cdot
\mathbf{z}_\perp) \Psi_{\pm}(\omega,z^2)/(2\pi)^2$, the normalized
B-meson WFs in the momentum space read as
\begin{equation}\label{waveWWa}
\tilde\Psi^{WW}_{+}(\omega,\mathbf{k}_\perp)=
\frac{\phi^{WW}_+(\omega)}{\pi}\delta\left(\mathbf{k}_\perp^2-
\omega(2\bar\Lambda-\omega)\right)
\end{equation}
and
\begin{equation}
\label{waveWWp} \tilde\Psi^{WW}_{-}(\omega,\mathbf{k}_\perp)=
\frac{\phi^{WW}_-(\omega)}{\pi}\delta\left(\mathbf{k}_\perp^2-
\omega(2\bar\Lambda-\omega)\right),
\end{equation}
whose $\mathbf{k}_\perp$-dependence correlates to the
$\omega$-dependence via a $\delta$-function.

Eqs.(\ref{waveWWa},\ref{waveWWp}) show that the WFs' dependence on
transverse and longitudinal momenta is strongly correlated through a
combined variable
$\mathbf{k}_\perp^2/[\omega(2\bar\Lambda-\omega)]$. Similar
transverse momentum behavior for the meson has been discussed in
Ref.\cite{bhl} by transforming the usual equal-time wavefunction to
its LC form, and has been derived in Ref.\cite{halperin} by adopting
the dispersion relations and the quark-hadron duality. In these two
references, the authors stated that the $k_T$-dependence of the
meson's wavefunction depends on the off-shell energy of the valence
quarks, i.e. $\sim \mathbf{k}_\perp^2/x(1-x)$, where $x$ is the
momentum fraction carried by the corresponding valence quark.

\subsection{Some constraints on the B-meson DAs $\phi_{\pm}(\omega)$}

The B-meson DAs can be obtained by taking the LC limit in
equations for the B-meson WFs. In the LC limit, Eq.(\ref{eq:1}) is
simplified as,
\begin{equation}
\omega \frac{d \phi_{-}(\omega)}{d \omega} + \phi_{+}(\omega) =
I(\omega) \label{phi1} \\
\end{equation}
with $I(\omega)=I(\omega,z^2)|_{z^2\to 0}$. Similarly, from
Eqs.(\ref{original2},\ref{original3}), we have
\begin{displaymath}
\frac{\partial\tilde\phi_+(t)}{\partial
t}+\frac{\partial\tilde\phi_-(t)}{\partial t}+
2i\bar\Lambda\tilde\phi_+(t)= -\int_0^1 du\;
2t[\tilde{\Psi}_A(t,u)+ \tilde{X}_A (t,u)+ 2u\,
\tilde{\Psi}_V(t,u) ]
\end{displaymath}
which leads to
\begin{equation}
\left(\omega - 2 \bar{\Lambda}\right)\phi_{+}(\omega) + \omega
\phi_{-}(\omega) = K(\omega) \label{phi2}
\end{equation}
with
\begin{displaymath}
K(\omega)= -2\frac{d}{d\omega} \int_{0}^{\omega}d\rho \int_{\omega -
\rho}^{\infty}\frac{d\xi}{\xi}  \left[ \Psi_{A}(\rho, \xi) +
X_{A}(\rho, \xi)\right] -4 \int_{0}^{\omega}d\rho \int_{\omega
-\rho}^{\infty} \frac{d\xi}{\xi}\frac{\partial \Psi_{V}(\rho,
\xi)}{\partial \xi}.
\end{displaymath}

The solution of the B-meson DAs can be conveniently decomposed
into two pieces as
\begin{equation}
\phi_{\pm}(\omega) = \phi_{\pm}^{WW}(\omega) +
\phi_{\pm}^{(g)}(\omega) \ ,\label{decomp}
\end{equation}
where $\phi_{\pm}^{WW}(\omega)$ are the DAs in the WW
approximation and $\phi_{\pm}^{(g)}(\omega)$ denote what induced
by the 3-particle source terms $I(\omega)$ and $K(\omega)$. From
Eqs.(\ref{phi1},\ref{phi2}), the solution for $\phi_{\pm}^{(g)}$
can be obtained straightforwardly, and reads:
\begin{equation}
\phi_{+}^{(g)}(\omega) = \frac{\omega}{2\bar{\Lambda}} {\cal
G}(\omega)+a_1\frac{K(\omega)} {\omega-2\bar\Lambda}\;, \;\;
\phi_{-}^{(g)}(\omega) = \frac{2\bar{\Lambda}
-\omega}{2\bar{\Lambda}}{\cal G}(\omega) +
a_2\frac{K(\omega)}{\omega}\ .\label{solpg}
\end{equation}
Here, $\omega \ge 0$, $a_1$ and $a_2$ are integration parameters
that satisfy the relation $a_{1} + a_{2}=1$. Note that the result
in Ref.\cite{qiao} is only a specific choice of $a_1=0$ and
$a_2=1$ in the general solution Eq.(\ref{solpg}). The function
${\cal G}(\omega)$ is expressed as follows:
\begin{eqnarray}
{\cal G}(\omega) &=& \theta(2\bar{\Lambda}-\omega)
\left\{\int_{0}^{\omega}d\rho \left[\frac{M(\rho)}{2\bar{\Lambda}
-\rho}+a_1\frac{K(\rho)}{(2\bar\Lambda-\rho)^2}\right]
-\frac{K(0)}{2\bar{\Lambda}}\right\}  \nonumber \\
&-& \theta(\omega - 2\bar{\Lambda}) \int_{\omega}^{\infty}d\rho
\left[\frac{M(\rho)}{2\bar{\Lambda} -\rho}+a_1\frac{K(\rho)}
{(2\bar\Lambda-\rho)^2}\right]-\int_{\omega}^{\infty}d\rho \left(
\frac{M(\rho)}{\rho} + a_2\frac{K(\rho)}{\rho^{2}} \right)\ ,
\end{eqnarray}
with $ M(\rho) = I(\rho) + \left( \frac{1}{2\bar{\Lambda}} -
a_2\frac{d}{d\rho}\right)K(\rho)$. One can easily check that
$\int_{0}^{\infty}d\omega \phi_{\pm}^{(g)}(\omega) = 0$, so the
total DAs are normalized \footnote{Strictly, it is not
true\cite{bdistribution1,bdistribution2}, however since the tail
of the LCDAs are proportional to $\alpha_s$, this could be
accepted as a "tree level" statement.}, i.e.
$\int_{0}^{\infty}d\omega \phi_{\pm}(\omega)= 1$.

As usual, we adopt the Mellin moments of $\phi_{\pm}(\omega)$,
which take the following form ($n= 0, 1, 2, \cdots$),
\begin{eqnarray}
\langle \omega^{n} \rangle_{\pm} = \int_{0}^{\infty} d\omega \
\omega^{n}\phi_{\pm}(\omega)&=&\int_{0}^{\infty} d\omega \
\omega^{n}\phi_{\pm}^{WW}(\omega) + \int_{0}^{\infty} d\omega \
\omega^{n}\phi_{\pm}^{(g)}(\omega)  \nonumber\\
&\equiv& \langle \omega^{n} \rangle_{\pm}^{WW} +\langle \omega^{n}
\rangle_{\pm}^{(g)}. \label{mel}
\end{eqnarray}
With the help of the formulae given in appendix B, we have
\begin{equation}
\langle \omega^{n} \rangle_{+}^{WW}=
\frac{2}{n+2}(2\bar{\Lambda})^{n} \ , \qquad \langle \omega^{n}
\rangle_{-}^{WW}= \frac{2}{(n+1)(n+2)}(2\bar{\Lambda})^{n} \ ,
\label{melsol}
\end{equation}
\begin{eqnarray}
\langle \omega^{n} \rangle_{+}^{(g)} &=&
\frac{2}{n+2}\sum_{i=1}^{n-1}(2\bar{\Lambda})^{i-1}
\sum_{j=1}^{n-i} {n-i \choose j} \left(
\left\{(n+1-i)\frac{2j+1}{j+1} + 1\right\}
\left[\Psi_{A}\right]^{n-i}_{j} \right. \nonumber\\
&+&\left. (n + 2 - i) \left[X_{A}\right]^{n-i}_{j} + (n + 3 -
i)\frac{j}{j+1} \left[\Psi_{V}\right]^{n-i}_{j} \right)
\label{melpg}
\end{eqnarray}
and
\begin{equation}
\langle \omega^{n} \rangle_{-}^{(g)} =\frac{1}{n+1}\langle
\omega^{n} \rangle_{+}^{(g)} - \frac{2n}{n+1}\sum_{j=1}^{n-1}{n-1
\choose j}\frac{j}{j+1} \left( \left[\Psi_{A}\right]^{n-1}_{j}
-\left[\Psi_{V}\right]^{n-1}_{j} \right) \ , \label{melmg}
\end{equation}
where ${i \choose j} = i!/[j!(i-j)!]$. The above results for
$\langle \omega^{n} \rangle_{\pm}^{(g)}$ have nothing to do with the
free parameters $a_1$ and $a_2$, due to the fact that $a_{1} +
a_{2}=1$, and hence they are in agreement with the ones obtained in
Ref. \cite{qiao}. It is obvious that the moments of DAs have no
relation to the 3-particle distribution $Y_A(\rho,\xi)$.

At the present, one knows little about the magnitudes of the
B-meson DAs' moments. In Ref.\cite{grozin}, the second moments of
the B-meson DAs are estimated by relating them to the matrix
elements of certain local operators and by calculating these
matrix elements from the sum rules in HQET, i.e.
\begin{equation}\label{sumlam}
\langle \omega^{2} \rangle_{+} = 2
\bar{\Lambda}^{2}+\frac{2}{3}\lambda_{E}^{2}
+\frac{1}{3}\lambda_{H}^{2} \ ,\quad\ \langle \omega^{2} \rangle_{-}
= \frac{2}{3} \bar{\Lambda}^{2} +\frac{1}{3}\lambda_{H}^{2} .
\end{equation}
Here $\lambda_{E}$ and $\lambda_{H}$ parameterize the matrix
elements of chromoelectric and chromomagnetic fields in the
$B$-meson rest frame,
\begin{equation}
\langle 0 |\bar{q} g \mbox{\boldmath $E$}\cdot\mbox{\boldmath
$\alpha$} \gamma_{5}h_{v} |\bar{B}(\mbox{\boldmath $p$}=0)\rangle =
f_{B}M \lambda_{E}^{2}\nonumber
\end{equation}
and
\begin{equation}
\langle 0 |\bar{q} g \mbox{\boldmath $H$}\cdot\mbox{\boldmath
$\sigma$} \gamma_{5}h_{v} |\bar{B}(\mbox{\boldmath $p$}=0)\rangle =
if_{B}M \lambda_{H}^{2} \ ,\nonumber
\end{equation}
with $E^{i}=G^{0i}$, $H^{i}=-\frac{1}{2}\epsilon^{ijk}G^{jk}$, and
$\mbox{\boldmath $\alpha$}= \gamma^{0}\mbox{\boldmath $\gamma$}$.
From Eqs.(\ref{mel}-\ref{sumlam}), we obtain
\begin{equation}\label{momentrelation1}
2\left[\Psi_{A}\right]_{1}^{1}+
\frac{3}{2}\left[X_{A}\right]^{1}_{1}+
\left[\Psi_{V}\right]^{1}_{1}
=\frac{2}{3}\lambda_{E}^{2}+\frac{1}{3}\lambda_{H}^{2}
\;,\;\;\left[\Psi_{V}\right]_{1}^{1}
+\frac{1}{2}\left[X_{A}\right]^{1}_{1}
=\frac{1}{3}\lambda_{H}^{2}\ .
\end{equation}
Together with Eq.(\ref{momentrelation2}), it leads to
\begin{equation}
\left[\Psi_{A}\right]_{1}^{1} = \frac{2}{3}\lambda_{E}^{2}\ ,
\quad \left[\Psi_{V}\right]_{1}^{1} =
\frac{1}{3}(\lambda_{H}^{2}+\lambda_{E}^{2})\ , \quad
\left[Y_{A}\right]^{1}_{1}=\left[X_{A}\right]^{1}_{1} =
-\frac{2}{3}\lambda_{E}^{2}.
\end{equation}
The above results are different from those in
Refs.\cite{qiao,alex}, where the contributions from
$\left[X_{A}\right]^{1}_{1}$ and $\left[Y_{A}\right]^{1}_{1}$ have
not been taken into consideration\footnote{According to the QCD
sum rule analysis \cite{alex}, as a rough estimation, the
contributions to the B-meson WF from $X_A$ and $Y_A$ are at least
suppressed by inverse power of the Borel parameter. Here we keep
both of them for a more complete estimation of the B-meson WFs.}.

As a summary, one may observe that $\phi_{\pm}(\omega)$ should
satisfy the following conditions:
\begin{eqnarray}
&&\int\phi_{-}(\omega)d\omega=1,\;\;
\int\phi_{+}(\omega)d\omega=1,\label{con1}\\
&&\int\omega\phi_{-}(\omega)d\omega=\frac{2}{3}\bar\Lambda,\;\;
\int\omega\phi_{+}(\omega)d\omega=\frac{4}{3}\bar\Lambda
,\label{con2}\\
&&\int\omega^2\phi_{-}(\omega)d\omega=
\frac{2}{3}\bar\Lambda^2+\frac{1}{3}\lambda_{H}^{2},\;\;
\int\omega^2\phi_{+}(\omega)d\omega=
2\bar\Lambda^2+\frac{2}{3}\lambda_{E}^{2}+
\frac{1}{3}\lambda_{H}^{2},\label{con3}
\end{eqnarray}
where according to the QCD sum rule analysis \cite{grozin},
$\lambda_{E}^2=0.11\pm 0.06$ and $\lambda_{H}^2=0.18\pm 0.07$.
Further more, as discussed in Refs. \cite{alex,braun}, the first
inverse moment of $\phi_{+}(\omega)$ should satisfy
\begin{equation}\label{con4}
\Lambda_0=\int \frac{d\omega} {\omega}\phi_{+}(\omega)=
\frac{1}{\lambda_B}.\;\;(\lambda_B=460\pm 160MeV)
\end{equation}

\subsection{Approximate solution for the B-meson WFs $\Psi_{\pm}(\omega,z^2)$
including 3-particle Fock states}

To solve the B-meson WFs $\Psi_{\pm}(\omega,z^2)$ including
3-particle Fock states, one need to know some more details on the
properties of the 3-particle WFs $\Psi_{V}(\rho,\xi,z^2)$,
$\Psi_{A}(\rho,\xi,z^2)$, $X_{A}(\rho,\xi,z^2)$ and
$Y_A(\rho,\xi,z^2)$, i.e. the transverse momentum dependence of
these WFs. In the following, we will take three assumptions so as to
provide an approximate solution for the B-meson WFs
$\Psi_{\pm}(\omega,z^2)$ from Eqs.(\ref{eq:1}-\ref{eq:4},
\ref{fr1},\ref{fr2}):

\noindent {\bf I}) Based on the B-meson WFs in the WW approximation
\cite{qiao,bwave}, we assume that $\Psi_+(\omega,z^2)$ and
$\Psi_-(\omega,z^2)$ have the same transverse momentum dependence,
i.e.
\begin{equation}\label{bwave}
\Psi_{\pm}[\omega, z^2]=\phi_{\pm}(\omega)\chi[\omega, z^2] ,
\end{equation}
and all the 3-particle WFs also have the same transverse momentum
dependence $\chi^{(h)}(\rho,\xi,z^2)$,
\begin{equation}\label{avwave}
\Psi_{A}(\rho,\xi,z^2)=\Psi_{A}(\rho,\xi)
\chi^{(h)}(\rho,\xi,z^2)\;,
\Psi_{V}(\rho,\xi,z^2)=\Psi_{V}(\rho,\xi) \chi^{(h)}(\rho,\xi,z^2)
\end{equation}
and
\begin{equation}
X_{A}(\rho,\xi,z^2)=X_{A}(\rho,\xi) \chi^{(h)}(\rho,\xi,z^2)\;,
Y_{A}(\rho,\xi,z^2)=Y_{A}(\rho,\xi)\chi^{(h)}(\rho,\xi,z^2),
\end{equation}
with the boundary condition $\lim_{z^2\to
0}\chi^{(h)}(\rho,\xi,z^2)=1$.

\noindent {\bf II}) Since the main features of the 3-particle DAs
are determined by its first several moments (the higher moments
will be suppressed by $\lambda_E$ or $\lambda_H$ accordingly), we
assume that the relation Eq.(\ref{momentrelation2}) among the
first non-zero double moments of the 3-particle DAs can be
extended to be a relation among the 3-particle DAs, i.e.
\begin{equation}\label{assum}
Y_A(\rho,\xi)\simeq\frac{X_A(\rho,\xi)- 3\Psi_A(\rho,\xi)}{4}\ .
\end{equation}

\noindent {\bf III}) We adopt a naive model \cite{alex} for the
difference between $\Psi_V(\rho,\xi)$ and $\Psi_A(\rho,\xi)$
\footnote{Even though the first moments of $\Psi_A(\rho,\xi)$ and
$\Psi_V(\rho,\xi)$ in Ref.\cite{alex} are different from our's, the
difference between these two moments are the same for both cases. So
we take the same model as the one in Ref.\cite{alex} for the
difference between $\Psi_V(\rho,\xi)$ and $\Psi_A(\rho,\xi)$.},
\begin{equation}\label{waveAV}
\Psi_V(\rho,\xi)-\Psi_A(\rho,\xi)=\frac{\lambda_{H}^2
-\lambda_{E}^2}{6\bar\Lambda^5}\rho\xi^2\exp
\left(-\frac{\rho+\xi}{\bar\Lambda}\right).
\end{equation}

With the help of Eqs.(\ref{fr1},\ref{fr2}) and the assumptions
(I,II), one can obtain two relations among the 3-particle WFs:
\begin{equation}\label{threerelation}
Y_A(\rho,\xi,z^2)=-\Psi_A(\rho,\xi,z^2)\;,\;\;\;
X_A(\rho,\xi,z^2)=-\Psi_A(\rho,\xi,z^2).
\end{equation}

\subsubsection{The transverse momentum dependence of $\Psi_{\pm}(\omega,z^2)$}

Applying Eq.(\ref{threerelation}) to Eqs.(\ref{eq:1}-\ref{eq:4}),
we obtain
\begin{eqnarray}\label{neq:1}
\omega \frac{\partial \Psi_{-}(\omega,z^2)}{\partial \omega} + z^2
\left(\frac{\partial \Psi_{+}(\omega,z^2)}{\partial z^2}
-\frac{\partial \Psi_{-}(\omega,z^2)}{\partial
z^2}\right)+\Psi_{+}(\omega,z^2) &=& I(\omega,z^2),\\
\label{neq:2} (\omega-2\bar\Lambda)\Psi_{+}(\omega,z^2)+
\omega\Psi_{-}(\omega,z^2) &=&-L(\omega,z^2),\\
\label{neq:3}
\bar\Lambda[\Psi_{+}(\omega,z^2)-\Psi_{-}(\omega,z^2)]
+2\frac{\partial^{2}}{\partial \omega\partial z^2}[
\Psi_{+}(\omega,z^2)+\Psi_{-}(\omega,z^2)]&=& L(\omega,z^2)\ ,
\end{eqnarray}
where Eq.(\ref{neq:2}) is obtained from the combination of
Eqs.(\ref{eq:2},\ref{eq:3}) and the source terms take the form,
\begin{eqnarray}
I(\omega,z^2) &=& 2\frac{d}{d\omega}\int_{0}^{\omega}d\rho
\int_{\omega - \rho}^{\infty}\frac{d\xi}{\xi}
\frac{\partial}{\partial \xi}\left[ \Psi_{A}(\rho, \xi,z^2)-
\Psi_{V}(\rho, \xi,z^2)\right]\nonumber\\
&& -2z^2\int_0^\omega d\rho\int_{\omega-\rho}^{\infty}
\frac{d\xi}{\xi}\left(\frac{\omega-\rho}{\xi}\right)
[\Psi_A(\rho,\xi,z^2)] \label{nfunI}
\end{eqnarray}
and
\begin{equation}
L(\omega,z^2)= 4\int_{0}^{\omega}d\rho \int_{\omega -
\rho}^{\infty}\frac{d\xi}{\xi}\frac{\partial}{\partial\xi}
[\Psi_V(\rho,\xi,z^2)]\label{nfunL}\ .
\end{equation}
Substituting Eqs.(\ref{bwave}, \ref{avwave}) into Eq.(\ref{neq:2}),
we obtain a relation between the transverse momentum distribution of
$\chi(\omega,z^2)$ and that of the 3-particle WFs,
\begin{equation}\label{tg0}
\chi(\omega,z^2)=\frac{\int_{0}^{\omega}d\rho \int_{\omega -
\rho}^{\infty}\frac{d\xi}{\xi}\frac{\partial}{\partial\xi}
[\Psi_{V}(\rho,\xi)\chi^{(h)}(\rho,\xi,z^2)]}{\int_{0}^{\omega}d\rho
\int_{\omega - \rho}^{\infty}\frac{d\xi}{\xi}
\frac{\partial}{\partial\xi} [\Psi_{V}(\rho,\xi)]}.
\end{equation}
It shows that if one knows the 3-particle WF
$\Psi_{V}(\rho,\xi,z^2)$ then the exact form of the transverse
momentum distribution $\chi(\omega,z^2)$ of the B-meson WFs can be
derived; and inversely, a constraint on $\Psi_{V}(\rho,\xi,z^2)$ can
be obtained as long as one knows the form of $\chi(\omega,z^2)$.

From Eqs.(\ref{neq:2}) and (\ref{neq:3}), we have
\begin{equation}\label{tg6}
2\frac{\partial^2}{\partial\omega\partial
z^2}f(\omega,z^2)=(\bar\Lambda-\omega)f(\omega,z^2)\ ,
\end{equation}
where $f(\omega,z^2)=[\Psi_+(\omega,z^2)+ \Psi_-(\omega,z^2)]$.
Eq.(\ref{tg6}) shows that the sum of the two WFs
$\Psi_{\pm}(\omega,z^2)$ do not explicitly depend on the
3-particle WFs. Based on the assumption (I), we rewrite
$f(\omega,z^2)$ as
\begin{equation}
f(\omega,z^2)=[\phi_+(\omega)+
\phi_-(\omega)]\chi\left(h(\omega)z^2\right)=
\kappa(\omega)\cdot\chi(x), \label{eq:55}
\end{equation}
where $x=\left[h(\omega)z^2\right]$ and the function $h(\omega)$
is to be determined. Substituting Eq.(\ref{eq:55}) into
(\ref{tg6}), we obtain
\begin{equation}\label{f-fun}
x \frac{d^2\chi(x)}{dx^2}+\frac{d\chi(x)}{dx}\left[1+
\frac{h(\omega)\kappa'(\omega)}
{h'(\omega)\kappa(\omega)}\right]+\chi(x)
\left[\frac{(\omega-\bar\Lambda)} {2h'(\omega)}\right]=0.
\end{equation}
To ensure the above equation for $\chi(x)$ be tenable to any value
of variable $\omega$, we set
\begin{equation}
\label{eq:57} \left[1+\frac{h(\omega)\kappa'(\omega)}
{h'(\omega)\kappa(\omega)}\right]=c_1\;,\;\;\left[\frac{(\omega-
\bar\Lambda)}{2h'(\omega)}\right]=c_2,
\end{equation}
where $c_1$ and $c_2$ are two parameters that are independent of
variable $\omega$. From Eq.(\ref{eq:57}) follows that
\begin{equation}\label{funh}
h(\omega)=\frac{\omega(\omega-2\bar\Lambda)}{4c_2}+c_3,
\end{equation}
where the value of $c_3$ is to be determined.

\begin{figure}
\centering
\includegraphics[width=0.55\textwidth]{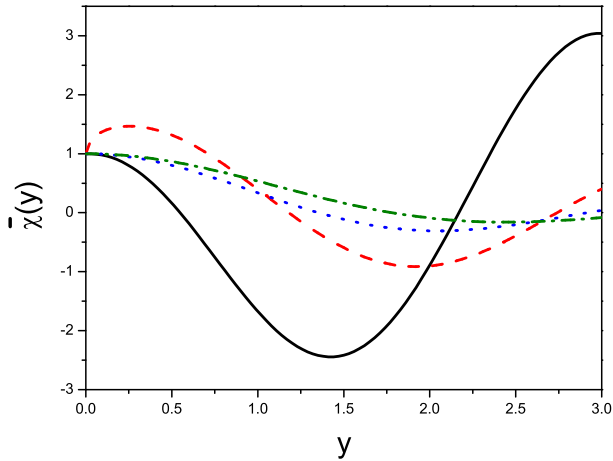}
\caption{Transverse distributions $\bar\chi(y)$ with different
values of $c_1$ and when $0<c_1\leq 1$, $K_2$ is fixed to be $1$.
The solid line, the dashed line, the dotted line and the dash-dot
line are for $c_1=0.2$, $0.8$, $1.2$ and $1.8$, respectively. }
\label{fig1}
\end{figure}

With the help of Eq.(\ref{eq:57}), the solution of
Eq.(\ref{f-fun}) can be generically expressed as
\begin{equation}\label{htrans}
\chi(x) = \bar\chi(y) = \left( K_1\Gamma[c_1] J_{c_{1}-1}[2y] +
K_2\Gamma[2-c_1]J_{1-c_{1}}[2y]
\right)y^{1-c_1}\;,\;\;(c_1\in(0,2))
\end{equation}
where $\Gamma$ is the usual Euler-gamma function, $J_n$ is the
modified Bessel $J$-functions, $K_{1,2}$ are undetermined constants
and $y=\sqrt{c_2 x}\;\;(c_2 x> 0)$. From the boundary condition
$\lim_{z^2\to 0}\chi(\omega,z^2)=1$, we obtain
\begin{equation}\label{eq77}
\lim_{y\to 0} \left( K_1\Gamma[c_1] J_{c_{1}-1}[2y] +
K_2\Gamma[2-c_1]J_{1-c_{1}}[2y]\right) y^{1-c_1}=1\ .
\end{equation}
Eq.(\ref{eq77}) leads to $K_1\equiv 1$ for any value of $c_1$, the
value of $K_2$ is arbitrary for $0<c_1<1$ and equals to $0$ for
$1\leq c_1<2$. Some typical distributions of $\bar\chi(y)$ are
shown in Fig.(\ref{fig1}).

Some discussions about the solution (\ref{htrans}) for the
transverse momentum dependence of the B-meson WFs are in the
following.

From the solutions in Eqs.(\ref{funh},~\ref{htrans}), one may find
that the WFs depend on $c_2$ and $c_3$ only in a combined form, so
the $(c_2c_3)$ in practice stands as one free parameter and
hereafter, we shall always replace $(4c_2c_3)$ by merely $c_3$ for
convenience.

When $c_1=1$ and $c_3=0$, we return to the transverse momentum
dependence of the B-meson wavefunction in the WW approximation.
The term including $c_3$ brings some differences to the
distribution under the WW approximation, especially the allowed
range for $\omega$ will be broadened for negative value of $c_3$.

\begin{figure}
\centering
\includegraphics[width=0.44\textwidth]{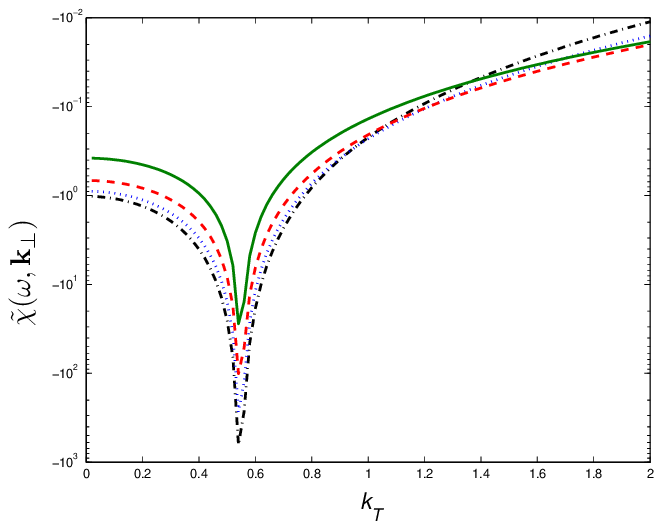}%
\hspace{0.3cm}
\includegraphics[width=0.44\textwidth]{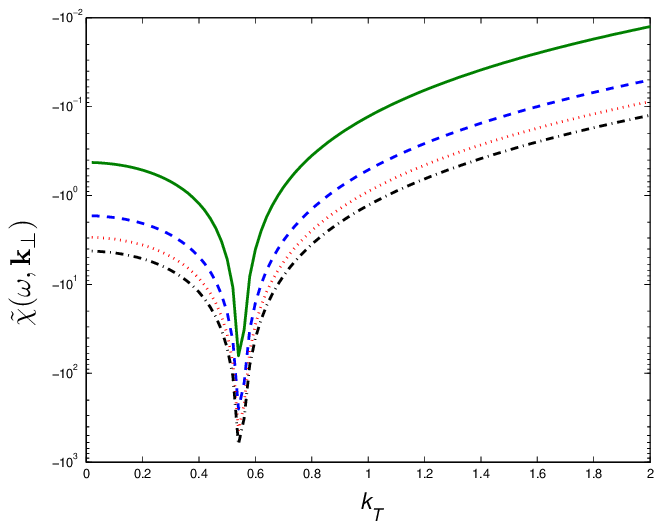}%
\caption{Transverse distributions of
$\tilde\chi(\omega,\mathbf{k}_\perp)$ ($k_T=|\mathbf{k}_\perp|$)
with fixed $\bar\Lambda=0.55GeV$ and $\omega=0.5GeV$. The left
diagram is drawn with fixed $K_2=1$ and with some different values
for $c_1$, i.e. $c_1=0.2$, $0.4$, $0.6$ and $0.8$, which are shown
in dash-dot, dotted, dashed and solid lines respectively. The
right diagram is drawn with fixed $c_1=0.6$ and different values
for $K_2$, i.e. $K_2=0$, $5.0$, $10.0$ and $15.0$, which are shown
in solid, dashed, dotted and dash-dot lines respectively.}
\label{fig2}
\end{figure}

When $c_1\geq 1$, one may find  the transverse momentum distribution
of the B-meson WF tends to be a $\delta$ function as the case in the
WW approximation. However, when $0<c_1<1$, the transverse momentum
distribution of the B-meson WF will be broadened. This is in
agreement with the conclusion drawn in Ref.\cite{qiao} that the
3-particle contributions might considerably broaden the B-meson
transverse momentum distribution. To show this point more clearly,
we take $c_3=0$ and transform the transverse part of
Eq.(\ref{htrans}) into the momentum space. For $c_1\in (0,1)$, we
obtain
\begin{equation}\label{momentum}
\tilde\chi(\omega,\mathbf{k}_\perp)=-\left(\frac{1}
{\Gamma[1-c_1]\pi}+\frac{\Gamma[2-c_1]\sin[\pi c_1]}{\pi^2} K_2
\right)\frac{\Gamma[2-c_1]}{{((2\bar\Lambda - \omega ) \omega) }
\left| 1 - \frac{k_T^2}{(2\bar\Lambda-\omega)\omega }
\right|^{2-c_1}},
\end{equation}
where $k_T=|\mathbf{k}_\perp|$. One may easily find that
$\tilde\chi(\omega,\mathbf{k}_\perp)$ satisfies the normalization
condition, $\int d^2\mathbf{k}_\perp\tilde\chi
(\omega,\mathbf{k}_\perp)=1$. The transverse distributions of
$\tilde\chi(\omega,\mathbf{k}_\perp)$ with fixed
$\bar\Lambda=0.55\; GeV$ and $\omega=0.5\; GeV$ are shown in
Fig.(\ref{fig2}). The left diagram of Fig.(\ref{fig2}) is drawn
with fixed $K_2=1$ and with some different choices for the
magnitude of $c_1$, i.e. $c_1=0.2$, $0.4$, $0.6$ and $0.8$. The
right diagram of Fig.(\ref{fig2}) is drawn with fixed $c_1(=0.6)$,
and varying $K_2$, i.e. $K_2=0$, $5.0$, $10.0$ and $15.0$. From
Fig.(\ref{fig2}), one may find that with the decreasing of $c_1$,
or increasing of $K_2$, the transverse momentum distributions
become broader and broader.

From Fig.(\ref{fig2}), one may observe that there is a dip in the
transverse momentum distributions around
$k_T=\sqrt{(2\bar\Lambda-\omega)\omega}$. It comes from the
denominator in Eq.(\ref{momentum}). Practically, such a dip will not
make any problem, due to the fact that we always need the integrated
results, which will be shown in Sec.IV. By taking the negative value
of $c_3$, the allowed range of $\omega$ will be broadened and it
will make a suppression to the singularity in Eq.(\ref{momentum}).
When summing up all the Fock states' contributions and taking the RG
evolution effects into consideration, one may expect a further
suppression to such singularity in the resultant transverse momentum
distributions \cite{bwaveevolute}.

\subsubsection{The distribution functions $\phi_{\pm}(\omega)$
with 3-particle Fock states}

The solutions for the distributions $\phi_{\pm}(\omega)$ with
3-particle Fock states have been given in
Eqs.(\ref{phiww},\ref{decomp},\ref{solpg}). However, since the
solutions for $\phi^{(g)}_{\pm}(\omega)$ (as shown in
Eq.(\ref{solpg})) involves the unknown 3-particle WFs, it can not
be used directly. In the following, we shall make an attempt to
provide more convenient expressions for $\phi_{\pm}(\omega)$ under
the above mentioned assumptions (I,II,III).

We can derive an expression for the sum of $\phi_{\pm}(\omega)$,
$\kappa(\omega)=\phi_{+}(\omega)+\phi_{-}(\omega)$, from
Eqs.(\ref{eq:57},\ref{funh}), which can be expanded in a more
convenient form as
\begin{equation}\label{comphi}
\kappa(\omega) =c_4 \times\exp
\left[\frac{2\bar\Lambda(c_{1}-1)}{(-c_{3})}
\omega\right]\times\left(1+\beta\omega\right),
\end{equation}
where $c_4$ is an overall normalization factor and we have
implicitly taken $(-c_3)>0$, which is reasonable, since the minus
sign indicates that the 3-particle WFs will broaden the allowed
range of $\omega$ in comparison to that in the WW approximation.
Here, $\beta$ is a new phenomenological parameter, which stands
for the summed effects of other expansion terms \footnote{those
terms in higher power series of $\omega$ can be summed up, since
when $\omega$ is big, their contributions are suppressed by the
overall exponential factor.}. Further more, the 3-particle source
term $I(\omega)$ can be simplified with the help of
Eqs.(\ref{funI},\ref{waveAV},\ref{threerelation}) as
\begin{equation}
I(\omega)= \frac{\lambda_{E}^2-\lambda_{H}^2}{18\bar\Lambda^6}
\omega(6\bar\Lambda^2-6\omega\bar\Lambda+\omega^2)\exp
\left(-\frac{\omega}{\bar\Lambda}\right).
\end{equation}
And then the solution for $\phi_{\pm}(\omega)$ can be obtained
from Eqs.(\ref{phi1},\ref{comphi}):
\begin{eqnarray}
\phi_{+}(\omega)&=&c_4 \beta\omega\times\exp
\left[\frac{2\bar\Lambda(c_{1}-1)}{(-c_{3})}
\omega\right]+c_4\omega\left(\beta+
\frac{2\bar\Lambda(c_{1}-1)}{(-c_{3})} \right){\rm
Ei}\left[\frac{2\bar\Lambda(c_{1}-1)}{(-c_{3})}
\omega\right]-K_{1}\omega\nonumber\\
&&-\left\{\frac{\lambda_{E}^2 -\lambda_{H}^2}{18\bar\Lambda^5}
\omega\exp \left(-\frac{\omega} {\bar\Lambda}\right)(-\omega+
5\bar\Lambda)+\frac{\lambda_{E}^2 -\lambda_{H}^2}{3\bar\Lambda^4}
\omega \; {\rm Ei}\left(-\frac{\omega}{\bar\Lambda}\right)\right\}
\label{phiplus}
\end{eqnarray}
and
\begin{eqnarray}
\phi_{-}(\omega)&=&c_4 \times\exp
\left[\frac{2\bar\Lambda(c_{1}-1)}{(-c_{3})}
\omega\right]-c_4\omega\left(\beta+
\frac{2\bar\Lambda(c_{1}-1)}{(-c_{3})} \right) {\rm Ei}
\left[\frac{2\bar\Lambda(c_{1}-1)}{(-c_{3})}
\omega\right]+K_{1}\omega\nonumber\\
&&+\left\{\frac{\lambda_{E}^2 -\lambda_{H}^2}{18\bar\Lambda^5}
\omega\exp \left(-\frac{\omega} {\bar\Lambda}\right)(-\omega+
5\bar\Lambda)+\frac{\lambda_{E}^2 -\lambda_{H}^2}{3\bar\Lambda^4}
\omega \;{\rm Ei}\left(-\frac{\omega}{\bar\Lambda}\right)\right\},
\label{phiminus}
\end{eqnarray}
where $K_1$ is an undetermined parameter and the exponential
integral function ${\rm Ei}(z)=-\int^{\infty}_{-z}e^{-t}/t dt$. All
the terms in the big parenthesis come from the source term
$I(\omega)$.

Some discussions about the solutions
(\ref{phiplus},\ref{phiminus}) for $\phi_{\pm}(\omega)$ are in the
following.

Eqs.(\ref{phiplus},\ref{phiminus}) show that $c_1$ and $c_3$ are
always in a combined form as $\left[\frac{2\bar\Lambda(c_{1}-1)}
{(-c_{3})}\right]$, so one can only get the combined results for
them. When $\omega\to 0$, we have $\phi_+(0)=0$ and
$\phi_-(0)=c_4$. Under the condition $c_3<0$, one may observe that
$c_1$ should be less than 1 to ensure that $\phi_{\pm}(\omega)$ is
normalizable, which shows that the transverse momentum dependence
of the B-meson wavefunction is broadened due to the introduction
of 3-particle wavefunctions.

From Eqs.(\ref{phi1}, \ref{comphi}), we obtain
\begin{displaymath}
\omega\phi_-(\omega)-2\int_0^\omega\phi_-(\rho)d\rho
=\int_0^\omega [I(\rho)-\kappa(\rho)]d\rho.
\end{displaymath}
Numerically, due to the fact that $\lambda_E^2 \sim \lambda_H^2$,
$\int_0^\omega I(\rho)d\rho << \int_0^\omega \kappa(\rho)d\rho$
for $\forall\;\omega>0$, and then one can safely set $I(\omega)=
0$, or equivalently $\lambda_E^2 \simeq\lambda_H^2$.

Under the approximation $\lambda_E^2 \simeq\lambda_H^2
\simeq2\bar\Lambda^2/3$, a solution for $\phi_{\pm}(\omega)$ can
be directly obtained by substituting
Eqs.(\ref{phiplus},\ref{phiminus}) into constraints
(\ref{con1},\ref{con2},\ref{con3}), i.e.
\begin{equation}\label{simplephi}
\phi_+(\omega)=\frac{\omega}{\omega_0^2}\exp \left(
-\frac{\omega}{\omega_0}\right),\;\;\;
\phi_-(\omega)=\frac{1}{\omega_0}\exp \left(
-\frac{\omega}{\omega_0}\right),
\end{equation}
where $\omega_0=2\bar\Lambda/3$. The undetermined parameters take
the following values,
$\left[\frac{2\bar\Lambda(c_{1}-1)}{(-c_{3})}\right]
=-\frac{1}{\omega_0}$, $\beta=\frac{1}{\omega_0}$, $K_1=0$ and
$c_4=\frac{1}{\omega_0}$. Such solution for $\phi_{\pm}(\omega)$
also satisfies the constraint (\ref{con4}), and it agrees well
with the model for $\phi_{\pm}(\omega)$ raised by
Ref.\cite{grozin}, where the same approximation $\lambda_E^2
\simeq\lambda_H^2 \simeq2\bar\Lambda^2/3$ is adopted in their QCD
sum rule analysis.

\section{A model for the B-meson WFs and its
phenomenological consequences}

In the above section, we have derived an approximate expression for
the B-meson WFs under the assumptions (I,II,III), in which the
3-particle Fock states' contributions are included. The transverse
momentum dependence of the B-meson WFs is shown in Eq.(\ref{htrans})
and the corresponding DAs are shown in
Eqs.(\ref{phiplus},\ref{phiminus}).

For the transverse momentum dependence of the 3-particle WFs, our
results indicate that when the value of $c_1$ is within the range
of $(0,1)$, it may be expanded to a hyperbola-like curve as shown
in Figs.(\ref{fig1},\ref{fig2}), rather than a simple
$\delta$-function as is the case of WW approximation. Our solution
for $\phi_{\pm}(\omega)$ favors $c_1<1$ under the condition that
$c_3<0$, which is reasonable since it means that the introduction
of 3-particle wavefunctions shall broaden the meson's longitudinal
and transverse distributions. This is in agreement with the
conclusion drawn in Ref.\cite{qiao}, where it has been argued that
the 3-particle contributions might considerably broaden the
B-meson transverse momentum distribution.

The solutions for the B-meson wavefunction in
Eqs.(\ref{htrans},\ref{phiplus},\ref{phiminus}) are somewhat
complex. Based on the discussions in Sec.III, we propose a simple
model for the B-meson wavefunction with 3-particle Fock states'
contributions in the following. For convenience, we write the two
normalized B-meson wavefunctions in the compact parameter $b$-space
(useful for the $k_T$-factorization approach \cite{kt}):
\begin{equation}\label{newmodel1}
\Psi_+(\omega,b)=\frac{\omega}{\omega_0^2}\exp \left(
-\frac{\omega}{\omega_0}\right) \Big(\Gamma[\delta] J_{\delta-1}
[\kappa] +(1-\delta)\Gamma[2-\delta] J_{1-\delta}[\kappa]\Big)\left(
\frac{\kappa}{2} \right)^{1-\delta}
\end{equation}
and
\begin{equation}
\Psi_-(\omega,b)=\frac{1}{\omega_0}\exp \left(
-\frac{\omega}{\omega_0}\right)\Big(\Gamma[\delta] J_{\delta-1}
[\kappa] +(1-\delta)\Gamma[2-\delta] J_{1-\delta}[\kappa]\Big)\left(
\frac{\kappa}{2} \right)^{1-\delta} ,\label{newmodel2}
\end{equation}
with $\omega_0=2\bar\Lambda/3$,
$\kappa=\theta(2\bar\Lambda-\omega) \sqrt{\omega
(2\bar\Lambda-\omega)}b $ and $\delta$ is in the range of $(0,1)$.
In the above model, $\lambda_E^2 \simeq\lambda_H^2
\simeq2\bar\Lambda^2/3$ is adopted, and to short the uncertainties
of the model as much as possible, we only take two main
phenomenological parameters $\bar\Lambda$ and $\delta$ into the
definition. Here for the transverse momentum dependence part, the
range of $\omega$ is fixed within the range of $(0,2\bar\Lambda)$
\footnote{This can be understood by checking the general solution
of $\phi_{\pm}(\omega)$ in Eqs.(\ref{phiplus},\ref{phiminus}),
i.e. the absolute value of $c_3$ can not be so big in order for
$\phi_{\pm}(\omega)$ to satisfy the constraints
(\ref{con1},\ref{con2},\ref{con3},\ref{con4}) under the condition
of $c_1<1$.}. When $\delta\to 1$, the transverse momentum
dependence of the B-meson wavefunction returns to that of the
B-meson wavefunction in the WW approximation. In the above
definition, the transverse momentum dependence of the B-meson
wavefunction is still the like-function of the off-shell energy of
the valence quarks and keeps the main features caused by the
3-particle Fock states, i.e. shall broaden the transverse momentum
dependence under the WW approximation to a certain degree.

In order to see the phenomenological influence of the 3-particle
Fock states' contributions, we recalculate the $B\to\pi$ transition
form factor within the $k_T$-factorization approach and show how
$\bar\Lambda$ and $\delta$ affect the final results. A consistent
analysis of the $B\to\pi$ transition form factor within its physical
range has been given in Ref.\cite{huangwu} by taking the WW B-meson
wavefunctions defined in Eqs.(\ref{waveWWa},\ref{waveWWp}). There we
shall adopt the same method as that of Ref.\cite{huangwu} to do the
calculations and to short the paper, we shall only list the results,
the interested reader may refer to Ref.\cite{huangwu} for more
details of the calculation technology.

Naively, the main property of the $B\to\pi$ transition form factor
is determined by the first inverse moment of $\phi_+(\omega)$, and
one may find from Eq.(\ref{con4}) that
$\left(\Lambda^{(g)}_0/\Lambda_0^{WW}\right)=
(\bar\Lambda-\lambda_B)/\lambda_B \sim 0.19$, where
$\Lambda^{(g)}_0$ is the first inverse moment of
$\phi^{(g)}_+(\omega)$ and $\Lambda_0^{WW}$ is that of
$\phi^{WW}_+(\omega)$. This shows that the 3-particle wavefunctions
might be small. In Ref.\cite{charng}, by studying the $B\to\gamma
l\nu$ decay within the perturbative QCD approach, the authors also
claims a small 3-particle contributions. More explicitly, in
Ref.\cite{charng}, the 3-particle contributions are estimated by
attaching an extra gluon to the internal off-shell quark line, and
then $(1/m_b)$ power suppression is readily induced. In the
following, we shall study the uncertainties caused by two parameters
$\delta$ and $\bar\Lambda$ under the condition that the 3-particle
contribution is less than $\pm 20\%$ of that of the WW case, and at
the same time give the possible range for $\delta$ and
$\bar\Lambda$.

\begin{figure}
\centering
\includegraphics[width=0.45\textwidth]{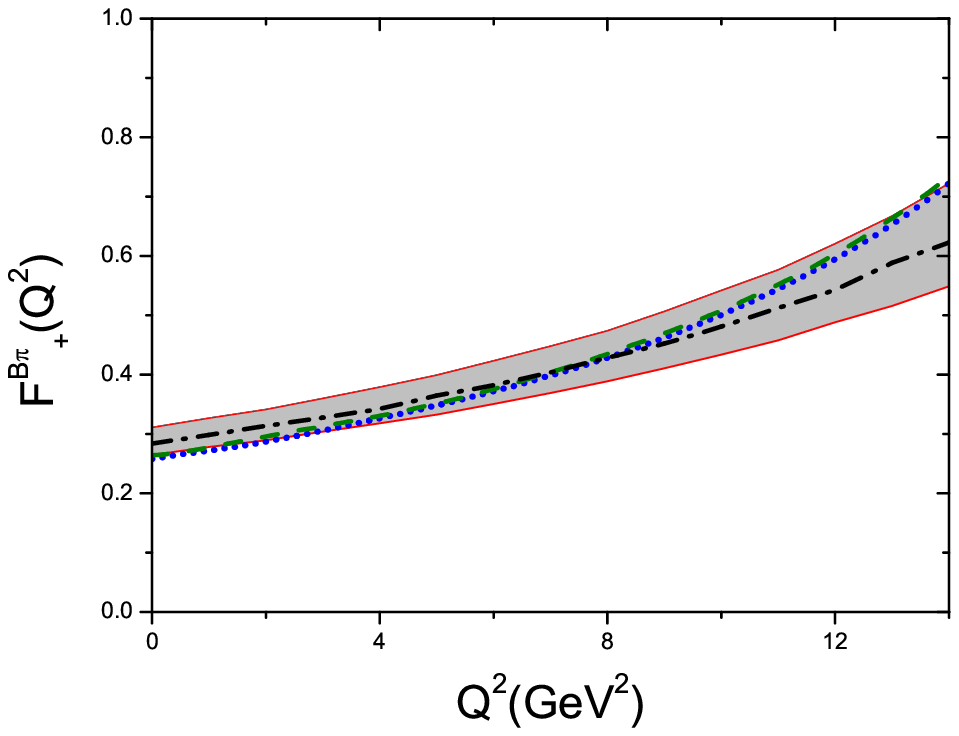}%
\hspace{0.2cm}
\includegraphics[width=0.45\textwidth]{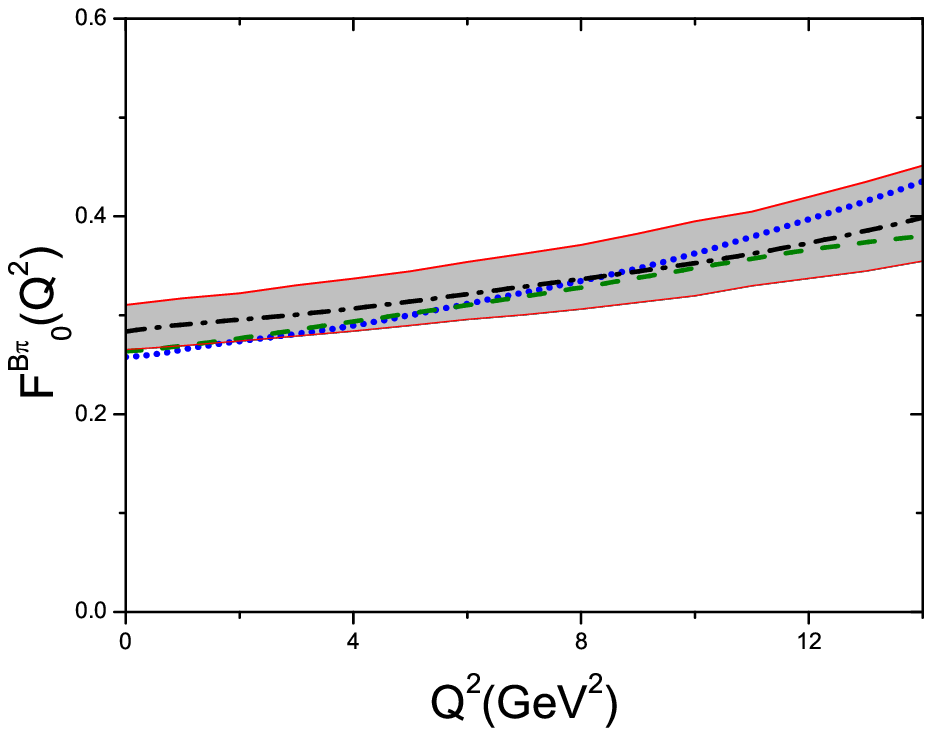}
\caption{B-meson transition form factors $F^{B\pi}_+(Q^2)$ (Left)
and $F^{B\pi}_0(Q^2)$ (Right) with the B-meson wavefunction
constructed in Eq.(\ref{newmodel1},\ref{newmodel2}). The upper
(lower) edge of the shaded band is for $\delta=0.30$
($\delta=0.25$), and the dash-dot line is for $\delta=0.27$. For
comparison, the dotted line and the dashed line are for the fitted
QCD light cone sum rule results \cite{sumrule} and the result
derived with WW B-meson wavefunction \cite{huangwu}, respectively.
$\bar\Lambda=0.55GeV$.} \label{fbpi1}
\end{figure}

\begin{figure}
\centering
\includegraphics[width=0.45\textwidth]{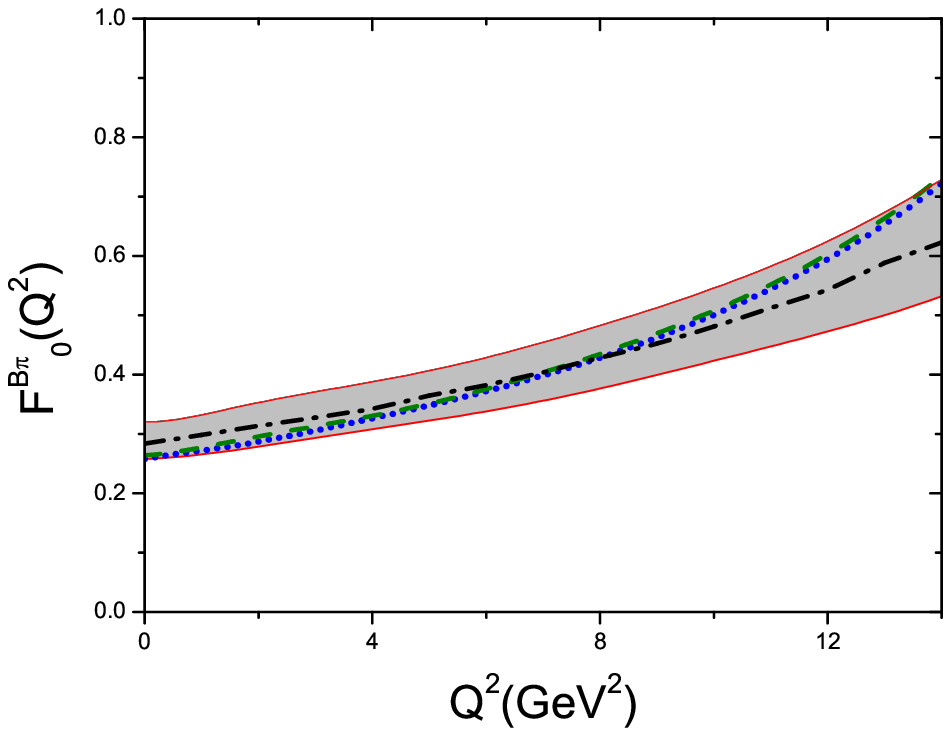}%
\hspace{0.2cm}
\includegraphics[width=0.45\textwidth]{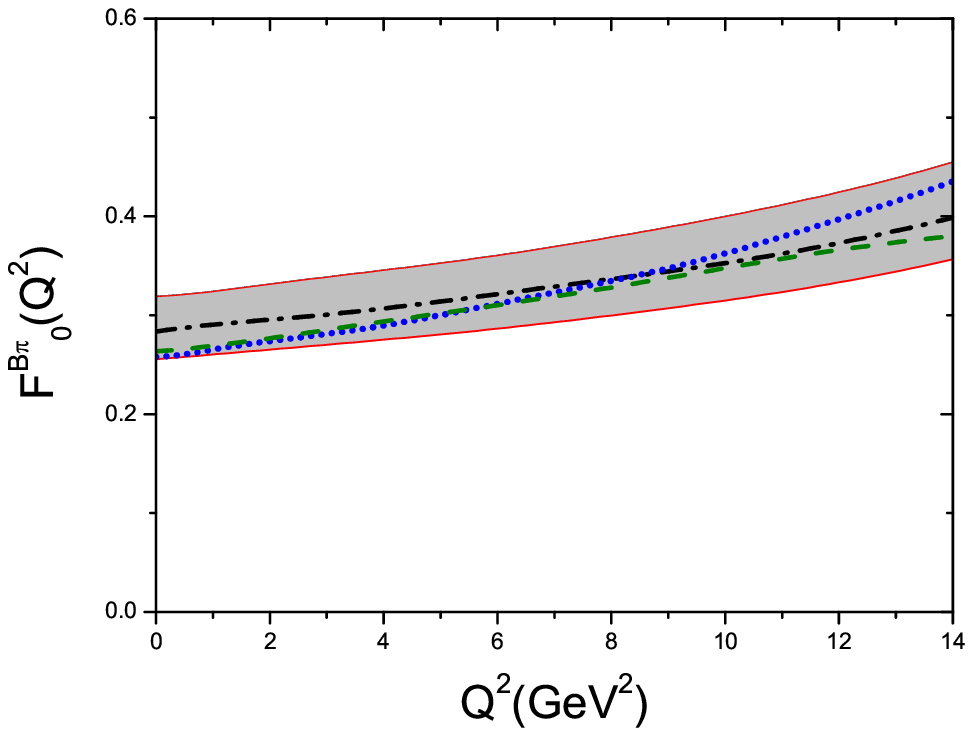}
\caption{B-meson transition form factors $F^{B\pi}_+(Q^2)$ (Left)
and $F^{B\pi}_0(Q^2)$ (Right) with the B-meson wavefunction
constructed in Eq.(\ref{newmodel1},\ref{newmodel2}). The upper
(lower) edge of the shaded band is for $\bar\Lambda=0.52$
($\bar\Lambda=0.58$), the dash-dot line is for $\bar\Lambda=0.55$.
For comparison, the dotted line and the dashed line are for the
fitted QCD light cone sum rule results \cite{sumrule} and the
result derived with WW B-meson wavefunction \cite{huangwu},
respectively. $\delta=0.27GeV$.} \label{fbpi2}
\end{figure}

By taking B-meson wavefunctions as
Eqs.(\ref{newmodel1},\ref{newmodel2}), we first study the
uncertainties of $B\to\pi$ transition form factor caused by
$\delta$ with fixed $\bar\Lambda=0.55GeV$ (the center value of
$\bar\Lambda$ determined in Ref.\cite{huangwu}). We show the
$B\to\pi$ transition form factors $F^{B\pi}_{+,0}(Q^2)$ in
Fig.(\ref{fbpi1}). Our results show that if the contribution from
the 3-particle wavefunction is limited to be within $\pm 20\%$ of
that of WW wavefunction with $Q^2\in (0,\sim 10GeV^2)$, then the
value of $\delta$ should be within the region of $(0.25,0.30)$.

Next, we study the uncertainties of $B\to\pi$ transition form
factor caused by $\bar\Lambda$ with fixed $\delta=0.27$ and the
results are shown in Fig.(\ref{fbpi2}). Our results show that if
the contribution from the 3-particle wavefunction is limited to be
within $\pm 20\%$ of that of WW wavefunction with $Q^2\in (0,\sim
10GeV^2)$, then $\bar\Lambda$ should be within the region of
$(0.52,0.58)GeV$. Similarly, one may find that if taking a larger
value for $\delta$ (e.g. $\delta=0.30$), then the range of
$\bar\Lambda$ should be shifted to a bigger interval (e.g.
$(0.55,0.61)GeV$).

Figs.(\ref{fbpi1},\ref{fbpi2}) show that by taking into account
the 3-particle wavefunctions' contributions, the $B\to\pi$
transition form factors $F^{B\pi}_{+,0}(Q^2)$ raise slower with
the increment of $Q^2$ than the case of WW B-meson wavefunction.
And if the contribution from the 3-particle wavefunction is
limited to be within $\pm 20\%$ of that of WW wavefunction with
$Q^2\in (0,\sim 10GeV^2)$, then the possible range of $\delta$ and
$\bar\Lambda$ are, $\delta\sim (0.25,0.30)$ and $\bar\Lambda\sim
(0.50GeV,0.60GeV)$.

\section{Summary}

It had been proved that the B-meson WF is renormalizable after
taking into account the RG evolution effects \cite{libwave}, and the
undesirable feature \cite{braun} of the B-meson DA can be removed
under evolution. Therefore, to keep the $k_T$ dependence in both the
hard scattering amplitude and the wavefunction is necessary. It was
found that the transverse and longitudinal momentum dependence in
the B-meson WF under the WW approximation is correlated through a
$\delta$-function,
$\delta(\mathbf{k}_\perp^2-\omega(2\bar\Lambda-\omega))$. In the
paper, we show that the transverse momentum distribution of the
B-meson WF can be broadened to be a hyperbola-like curve by
including 3-particle Fock state, rather than a simple
$\delta$-function.

The solutions in this paper provide a practical framework for
constructing the B-meson LC WFs $\Psi_{\pm}(\omega,z^2)$ and hence
are meaningful for phenomenological applications. And we have
constructed a new model for the B-meson wavefunction in the compact
parameter $b$-space as shown in
Eqs.(\ref{newmodel1},\ref{newmodel2}) based on these solutions.
There are uncertainties caused by two unknown parameters
$\bar\Lambda$ and $\delta$. However, since the B-meson WFs are
universal, we can determine them by global fitting of the
experimental data. By taking $B\to\pi$ transition form factor as an
example, we show that if the 3-particle wavefunctions' contributions
are less than $20\%$ of that of the WW case, then one may observe
that the preferable values for these two parameters are $\delta\sim
0.27$ and $\bar\Lambda\sim 0.55GeV$.

The reasonable inclusion of the 3-particle Fock states in B-meson
WFs provides us with the chance to make a more precise evaluation
on the B meson decays. Further studies on the B-meson WFs with
higher Fock states and its phenomenological implications are still
necessary.

\begin{center}
\section*{Acknowledgements}
\end{center}

This work was supported in part by the Natural Science Foundation
of China (NSFC). C.-F. Qiao thanks the Institute for Nuclear
Theory at the University of Washington for its hospitality and the
Department of Energy for partial support during the completion of
this work. X.-G. Wu thanks the support from the
China Postdoctoral Science Function. \\

\appendix
\section{Basic formulae for the Fourier transformation}

First, we define $\tilde{F}(t, u)=\tilde{F}(t,u,z^2)|_{z^2\to 0}$
and the Fourier transformation:
\begin{eqnarray}
\tilde{\Psi}_{\pm}(t, z^{2}) &=& \int d\omega \ e^{-i \omega t}
\Psi_{\pm}(\omega, z^{2})\ ,\\
\tilde{F}(t, u) &=& \int d\omega d \xi \ e^{-i(\omega  + \xi u)t}
F(\omega, \xi).\qquad (F=\{ \Psi_{V}, \Psi_{A}, X_{A}, Y_{A} \})
\end{eqnarray}
Here $\omega v^{+}$ and $\xi v^{+}$ ($v^+$ is the `$+$'-component of
$v$ defined in Eq.(\ref{hqeteq})) denote the LC projection of the
momentum carried by the light antiquark and the gluon, respectively,
and $F(\omega, \xi)$ vanishes unless $\omega \ge 0$ and $\xi \ge 0$.

Some useful formulae:
\begin{eqnarray}
\int \frac{dt}{2\pi} e^{i\omega
t}\tilde\Psi(t,z^2)&=&\Psi(\omega,z^2),\\
\int \frac{dt}{2\pi} e^{i\omega
t}\frac{\partial\tilde\Psi(t,z^2)}{\partial
t}&=&-i\omega\Psi(\omega,z^2),\\
\int \frac{dt}{2\pi} t^n e^{i\omega t}\tilde\Psi(t,z^2)
&=&(-i)^n\frac{\partial^n}{\partial\omega^n}\Psi(\omega,z^2),\qquad
(n\geq 1)\\
\int \frac{dt}{2\pi} t^n e^{i\omega
t}\frac{\partial\tilde\Psi(t,z^2)}{\partial
t}&=&(-i)^n\frac{\partial^n}{\partial\omega^n}[-i\omega\Psi(\omega,z^2)],\qquad
(n\geq 1)
\end{eqnarray}
where $\Psi=(\Psi_+,\Psi_-)$ and
$\tilde\Psi=(\tilde\Psi_+,\tilde\Psi_-)$, respectively. And for
the 3-particle distributions, we have
\begin{eqnarray}
\int\frac{du dt}{2\pi}e^{i\omega t}\tilde\Psi(t,u)&=&\int_0^\omega
d\rho\int_{\omega-\rho}^{\infty}\frac{d\xi}{\xi}\Psi(\rho,\xi),\\
\int\frac{du dt}{2\pi}t^n e^{i\omega
t}\tilde\Psi(t,u)&=&(-i)^n\frac{\partial^n}{\partial\omega^n}\int_0^\omega
d\rho\int_{\omega-\rho}^{\infty}\frac{d\xi}{\xi}\Psi(\rho,\xi),\qquad
(n\geq 1)\\
\int\frac{u du dt}{2\pi}e^{i\omega
t}\tilde\Psi(t,u)&=&\int_0^\omega d\rho\int_{\omega-\rho}^{\infty}
\frac{d\xi}{\xi}\left(\frac{\omega-\rho}{\xi}\right)
\Psi(\rho,\xi),\\
\int\frac{ut du dt}{2\pi}e^{i\omega
t}\tilde\Psi(t,u)&=&(-i)\int_0^\omega
d\rho\int_{\omega-\rho}^{\infty} \frac{d\xi}{\xi}\frac{\partial
\Psi(\rho,\xi)}{\partial\xi},\\
\int\frac{ut^n du dt}{2\pi}e^{i\omega
t}\tilde\Psi(t,u)&=&(-i)^n\frac{\partial^{n-1}}{\partial\omega^{n-1}}\int_0^\omega
d\rho\int_{\omega-\rho}^{\infty} \frac{d\xi}{\xi}\frac{\partial
\Psi(\rho,\xi)}{\partial\xi},\qquad (n\geq 2)
\end{eqnarray}
where $\Psi=(\Psi_{V}, \Psi_{A}, X_{A}, Y_{A})$ and
$\tilde\Psi=(\tilde\Psi_{V}, \tilde\Psi_{A}, \tilde X_{A}, \tilde
Y_{A})$, respectively. Note here, we have implicitly using the
following equation,
\begin{equation}
\int_0^{\infty}d\rho\int_0^{\infty}d\xi\int_0^1 du
\Psi(\rho,\xi)\cdot \delta(\omega-\rho-\xi u)=\int_0^\omega
d\rho\int_{\omega-\rho}^{\infty}\frac{d\xi}{\xi}\Psi(\rho,\xi).
\end{equation}

\section{Mellin moments of the distribution amplitude}

In order to calculate the Mellin moments defined in
Eq.(\ref{mel}), it is more convenient to use the derivative of
${\cal G}(\omega)$, i.e.
\begin{equation}
\frac{d}{d\omega}{\cal G}(\omega)=\frac{2\bar\Lambda}
{\omega(2\bar\Lambda-\omega)}\left[I(\omega)+\frac{a_1
K(\omega)}{2\bar\Lambda-\omega}+\frac{a_2
K(\omega)}{\omega}-a_2\frac{d}{d\omega}K(\omega)\right].
\end{equation}
And from Eq.(\ref{phi1}), we have the following equation,
\begin{equation}\label{momentphi}
\langle \omega^{n} \rangle_{-}=\frac{\langle\omega^{n}\rangle_{+}}
{n+1}-\frac{1}{n+1}\int_0^{\infty}\omega^nI(\omega)d\omega.
\end{equation}
Substituting Eq.(\ref{funI}) into Eq.(\ref{momentphi}), we get
\begin{eqnarray}
\int_0^{\infty}\omega^nI(\omega)d\omega
&=&2\int_0^{\infty}\omega^n\frac{d}{d\omega} \left[
\int_0^{\infty}d\rho\int_0^{\infty}d\xi\int_0^1
du\frac{\partial}{\partial\xi}[\Psi_A-\Psi_V]\cdot
\delta(\omega-\rho-\xi u)\right]d\omega\nonumber\\
&=&-2n\int_0^{\infty}d\rho\int_0^{\infty}d\xi\int_0^1 du(\rho+\xi
u)^{n-1} \frac{\partial}{\partial\xi}[\Psi_A-\Psi_V]\nonumber\\
&=& 2n\int_0^{\infty}d\rho\int_0^{\infty}d\xi
\sum_{j=2}^n\frac{j-1}{n}{n \choose n-j}\rho^{n-j}\xi^{j-2}
[\Psi_A-\Psi_V]\nonumber\\
&=& 2n\sum_{j=1}^{n-1}\frac{j}{j+1}{n-1 \choose
j}[\Psi_A-\Psi_V]^{n-1}_j \ ,
\end{eqnarray}
where the double moments of the 3-particle distributions are
defined as
\begin{equation}
\left[F\right]^{i}_{j} =\int_{0}^{\infty}d\rho
\int_{0}^{\infty}d\xi \rho^{i-j} \xi^{j-1} F(\rho, \xi) \qquad
(F=\{ \Psi_{V}, \Psi_{A}, X_{A}\})\ .
\end{equation}
The moments of $\phi^{(g)}_+(\omega)$ can be written as
\begin{eqnarray}
\langle\omega^{n}\rangle^{(g)}_{+}&=&-\frac{1}{n+2}\int_0^\infty
\frac{\omega^{n+1}}{2\bar\Lambda-\omega}\left[I(\omega)+\frac{a_1
K(\omega)}{2\bar\Lambda-\omega}+\frac{a_2 K(\omega)}
{\omega}-a_2\frac{d}{d\omega}K(\omega)\right]d\omega\nonumber\\
&& +a_1\int_0^\infty\frac{\omega^n}{\omega-2\bar\Lambda}K(\omega)
d\omega \ .
\end{eqnarray}
More definitely, with the help of the Eqs.(\ref{funI},\ref{funK}),
we have
\begin{eqnarray}
\langle\omega^{n}\rangle^{(g)}_{+}&=& -\frac{2}{n+2}
\int\frac{\omega^{n+1}}{2\bar\Lambda-\omega}\frac{d}{d\omega}\left[
\int \int\int d\rho d\xi du\frac{\partial}{\partial\xi}[\Psi_{A}-
\Psi_{V}]\cdot \delta(\omega-\rho-\xi u)\right]d\omega\nonumber\\
&+&\frac{2}{n+2}\int\frac{\omega^{n+1}}
{(2\bar\Lambda-\omega)^2}\left\{\frac{d}{d\omega}\left[ \int
\int\int d\rho d\xi du[\Psi_{A}+X_{A}]\cdot \delta(\omega-\rho-\xi
u)\right]\right.\nonumber \\
&& \left. \qquad\qquad\qquad\qquad+2\left[\int \int\int d\rho d\xi
du\frac{\partial}{\partial\xi}[\Psi_{V}]\cdot
\delta(\omega-\rho-\xi u)\right] \right\}d\omega\nonumber\\
&+&2\int\frac{\omega^{n}}
{2\bar\Lambda-\omega}\left\{\frac{d}{d\omega}\left[ \int \int\int
d\rho d\xi du[\Psi_{A}+X_{A}]\cdot \delta(\omega-\rho-\xi
u)\right]\right.\nonumber \\
&& \left. \qquad\qquad\qquad\qquad+2\left[\int \int\int d\rho d\xi
du\frac{\partial}{\partial\xi}[\Psi_{V}]\cdot
\delta(\omega-\rho-\xi u)\right] \right\}d\omega\ ,\nonumber
\end{eqnarray}
where we have implicitly applied the relation $a_1+a_2=1$. To get
the final results, the following formulae are useful $(n\geq 2)$,
\begin{eqnarray}
\int_0^{\infty}\omega^n\frac{d}{d\omega} \left[
\int_0^{\infty}d\rho\int_0^{\infty}d\xi\int_0^1
du\frac{\partial}{\partial\xi}[F]\cdot \delta(\omega-\rho-\xi
u)\right]d\omega &=& n\sum_{j=1}^{n-1}\frac{j}{j+1}{n-1 \choose
j}[F]^{n-1}_j\ ,\nonumber\\
\int_0^{\infty}\omega^n\frac{d}{d\omega} \left[
\int_0^{\infty}d\rho\int_0^{\infty}d\xi\int_0^1 du[F]\cdot
\delta(\omega-\rho-\xi u)\right]d\omega &=& -\sum_{j=1}^{n}{n
\choose j}[F]^{n}_j\ ,\nonumber\\
\int_0^{\infty}\omega^n\left[
\int_0^{\infty}d\rho\int_0^{\infty}d\xi\int_0^1
du\frac{\partial}{\partial\xi}[F]\cdot \delta(\omega-\rho-\xi
u)\right]d\omega &=& -\sum_{j=1}^{n}\frac{j}{j+1}{n \choose
j}[F]^{n}_j\ ,\nonumber\\
\int_0^{\infty}\omega^n\frac{d^2}{d\omega^2} \left[
\int_0^{\infty}d\rho\int_0^{\infty}d\xi\int_0^1 du[F]\cdot
\delta(\omega-\rho-\xi u)\right]d\omega &=& n\sum_{j=1}^{n-1}{n-1
\choose j}[F]^{n-1}_j\ ,\nonumber
\end{eqnarray}
where $F=(\Psi_{V}, \Psi_{A}, X_{A}, Y_{A})$.

With the help of the above formulae, the value of
$\langle\omega^{n}\rangle^{(g)}_{+}$ can be directly derived, as
is shown in Eq.(\ref{melpg}). One subtle point is that, before
doing the integration over $\omega$, it is more useful to expand
$\frac{1}{2\bar\Lambda-\omega}$, i.e.
\begin{equation}
\frac{1}{2\bar\Lambda-\omega}=\frac{-1}{\omega}\left(1
+\left(\frac{2\bar\Lambda}{\omega} \right)
+\left(\frac{2\bar\Lambda}{\omega}\right)^2 +
\left(\frac{2\bar\Lambda}{\omega}\right)^{3}+\cdots\right),
\end{equation}
where the power of $\left(\frac{2\bar\Lambda}{\omega}\right)$ will
be stopped at a particular value, for one may observe that the
terms with even higher powers contribute zero exactly. And to do
the integration in a form like
\begin{equation}
\int \frac{\omega^n}{(\omega-2\bar\Lambda)^2}\frac{d {\cal
H(\omega)}}{d\omega} d\omega,
\end{equation}
where ${\cal H(\omega)}$ is a function of $\omega$, we can
transform it to a more familiar one
\begin{equation}
\int \frac{\omega^n}{(\omega-2\bar\Lambda)^2}\frac{d {\cal
H(\omega) }}{d\omega} d\omega=n\int \frac{
\omega^{n-1}}{(\omega-2\bar\Lambda)}\frac{d{\cal H(\omega)
}}{d\omega} d\omega+\int \frac{\omega^n}{(\omega-2\bar\Lambda)}
\frac{d^2 {\cal H(\omega) }}{d\omega^2} d\omega\ .
\end{equation}

\end{document}